\documentclass{llncs}
\usepackage[utf8x]{inputenc}
\usepackage{url}
\begin{document}

\title{Solving XCSP problems by using Gecode}


\author{Massimo Morara \and Jacopo Mauro \and Maurizio Gabbrielli}

\institute{University of Bologna.\\
	\email{ morara | jmauro | gabbri@cs.unibo.it}
}

\maketitle

\begin{abstract}
Gecode is one of the most efficient libraries that can be used for constraint solving. However, using it requires dealing with C++ programming details. On the other hand several formats for representing constraint networks have been proposed. Among them, XCSP has been proposed as a format based on XML which allows us to represent constraints defined either extensionally or intensionally, permits global constraints and has been the standard format of the international competition of constraint satisfaction problems solvers.
In this paper we present a plug-in for solving problems specified in XCSP by exploiting the Gecode solver. This is done by dynamically translating constraints into Gecode library calls, thus avoiding the need to interact with C++.

\end{abstract}

\section{Introduction}

Constraint Programming \cite{Handbook_of_Constraint_Programming} has attracted high attention among experts from many areas because of its potential for solving hard real life problems and because it is based on a strong theoretical foundation. The success of Constraint Programming (CP) derives from the fact that on one hand it allows to model a problem in a simple way and on the other hand it provides an efficient problem solving algorithms.
However, the CP community lacks a standardized representation of problem instances and this still limits the acceptance of CP by the business world. One attempt to overcome this problem was taken by the Association for Constraint Programming with the proposal of Java Specification Request JSR-331 ``Constraint Programming API" \cite{standardsSite,JSR331Site}. The goal of this specification is the creation of a powerful API for specifying CP problems. In the last five years other approaches focusing on more low level languages emerged. The aim of these approaches is to define a minimal domain dependent language that supports all the major constraint features and requires, at the same time, a minimal implementation effort to be supported by constraint solvers. Two languages following this goal are worth mentioning: FlatZinc \cite{FlatZinc} and XCSP \cite{XCSP}. The former was originally created to be the target language into which a higher level CSP instance (e.g. a CSP modeled with MiniZinc \cite{DBLP:conf/cp/NethercoteSBBDT07}) is translated. Today FlatZinc is also used as a low level ``lingua franca" for solver evaluation and testing. For instance, since 2008 FlatZinc has been used in the MiniZinc Challenge \cite{MiniZincChallange,DBLP:journals/constraints/StuckeyBF10}, a competition where different solvers are compared by using a benchmark of MiniZinc instances that are compiled into FlatZinc.

XCSP is a language structurally very similar to FlatZinc. XCSP was defined with the purpose of being a unique constraint model that could be used by all the CP solvers. It was first proposed in 2005 for the solvers competing in the International CSP Solver Competition \cite{CSPCompetition}, and has then been used in other contexts and extended. In this paper, we focus on XCSP. In particular, we consider its current version, i.e. XCSP version 2.1.

The need of a standard is also caused by the huge number and diversity of solvers. Today only few solvers support natively FlatZinc or XCSP. Unfortunately, Gecode \cite{Gecode}, one of the most well known and used solvers, support FlatZinc only.
For this reason we have created \emph{x4g}, a plug-in that allows us to solve problems defined in XCSP by using the Gecode solver. The goal of x4g is twofold. Firstly, we want to exploit Gecode for solving problems that are specified in XCSP without considering low level implementation details or writing a single line of code in C++. Secondly, we want to provide a tool that can be used to evaluate the performances of the Gecode solver with respect to the other entries of the  International Solver Competition. This could be very interesting since, to the best of our knowledge, the benchmark used in the International Solver Competition is the biggest available to the CP community.\footnote{The MiniZinc Challenge has a smaller benchmark of instances and fewer participants than the International Solver Competition.}

In the reset of this paper, we present in Section \ref{overview} a brief overview of the XCSP language and of the Gecode solver. In Section \ref{x4g} we describe the idea behind the x4g plug-in. Section \ref{conclusions} concludes by mentioning some directions for future works.

\section{Background}
\label{overview}
An extensive presentation of the XCSP format and the Gecode solver is beyond the scope of this paper. Here we just give a short overview of XCSP and Gecode.

\subsection{XCSP}
XSCP is an extended format to represent constraint networks using XML. The Extensible Markup Language (XML) is a simple and flexible text format playing an increasingly important role in the exchange of a wide variety of data on the Web. The objective of the XML representation (in XCSP) is to ease the effort required to test and compare different algorithms by providing a common testbed of constraint satisfaction
instances.
The proposed representation is low-level: for each instance the domains, variables, relations (if any), predicates (if
any) and constraints are exhaustively defined. No control flow constructs like``for" cycles or ``if then else" statements can be used.

Roughly speaking, there exist two variants of this format: a fully-tagged representation and an abridged representation. The first one is a full XML, completely structured representation which is suitable for using generic XML tools but is quite verbose and tedious to use for a human being.
The second representation is just a shorthand notation of the first one and it is easier to read and to write for a human being, but less suitable for generic XML tools.

As an example of an XCSP program consider the following one where the well known ``all different" constraint is applied to two variables  $A1$ and $A2$ which can assume values only in the domain $[1,2]$.
  \begin{verbatim}
<domains nbDomains="1">
  <domain name="d0" nbValues="2">1..2</domain>
</domains>
<variables nbVariables="2">
  <variable name="A1" domain="d0"/>
  <variable name="A2" domain="d0"/>
</variables>
<constraint name="c0"  arity="2"
     scope="A1 A2"
     reference="global:alldifferent"/>
\end{verbatim}

\subsection{Gecode}
Gecode (Generic Constraint Development Environment) provides a constraint solver with state-of-the-art performance while being modular and extensible. It supports the programming of new propagators, branching strategies, and search engines. New domains can be programmed at the same level of efficiency as finite domain and integer set variables that are already predefined. Furthermore Gecode is  distributed under a very permissive license, it is portable and well documented, and comes with a complete tutorial. All these features have made Gecode one of the preferred choices for solving CSPs.

Gecode is written in C++ and supports the FlatZinc format through an external plug-in that is able to parse a FlatZinc instance and solve it using Gecode library calls. The use of this plug-in allowed Gecode to participate to the 2010 MiniZinc challenge \cite{MiniZincChallange}, where it won all the tracks of the competition.

\section{x4g}
\label{x4g}

In principle the translation of XCSP instances into Gecode is similar to the task performed by the Gecode/FlatZinc plug-in \cite{GecodeFlatZinc} that parses a FlatZinc instance and produces an internal data structure (Gecode Space Object) that can be later used to retrieve the solution of the CSP problem.

To use Gecode in order to solve CSPs defined in XCSP, one could try implement a XCSP to FlatZinc compiler. However, in order to avoid potential loss of information, to be more flexible and less dependent on the Gecode/FlatZinc plug-in, we chose to provide a direct translation from XCSP into Gecode. Hence we have developed x4g; a plug-in that parses an XCSP instance and, for every constraint defined within the XCSP file, generates an equivalent number of Gecode constraints. When all the XCSP constraints are translated into Gecode constraints, a Gecode Space Object is returned. This object can later be used to get a solution to the CSP problem by using one of the many predefined search strategies provided by Gecode or local search strategies following \cite{DBLP:conf/hm/CiprianoGD09}.\footnote{Note that the XCSP format specifies only the constraints, the choice of the search strategy is left to the user.}

To develop the x4g plug-in we used the XCSP parser provided by the International CSP competition organizers. This parser, developed in particular to support the abridged notation, is written in C++  by using the well known libxml2 libraries \cite{libxml2}. Unfortunately, the parser supports only a limited number of global constraints. Therefore, we modified it in order to support additional ones. 

The XCSP format is mainly used to specify CSPs but it also supports extensions to define weighted constraints or quantifiers over constraints. x4g was designed to target only CSPs, hence it does not support these additional features.

XCSP supports only finite domains. Since Gecode supports finite domains too, the domain encoding from XCSP to Gecode was straightforward.
XCSP provides a construct which allows one to define relations over variables. These could be seen as constraints listing all the admissible values that some variables can take. XCSP relations are mapped into Gecode by using extensional constraints. In XSCP the semantics of a relation can also be given by stating all the non admissible values of the variables (i.e. conflicts between variables). We used inequality constraints for translating these relations into Gecode.
Another construct of XCSP is predicate, a boolean parametric expression that is considered satisfied if and only if it evaluates to true. The number of parameters in a predicate is fixed and, differently from relations, predicates allow integer parameters. The mapping of the predicates was also straightforward because, with only few exceptions, Gecode has for every predicate an API for posting an equivalent constraint.
As far as global constraints are concerned, XCSP supports the majority of the global constraints defined in the Global Constraint Catalog \cite{GlobalConstraintCatalog}. Since in this catalog there are hundreds of global constraints, a full XCSP support means to provide an encoding for a huge number of global constraints. This was out of the scope of the project, we have just chosen to support a subset of the most used global constraints. Currently x4g supports the following global constraints: {\tt alldifferent}, {\tt among}, {\tt atleast}, {\tt atmost}, {\tt cumulative}, {\tt diffn}, {\tt disjunctive}, {\tt element}, {\tt global\_cardinality}, {\tt lex\_less}, {\tt lex\_lesseq}, 
\\
{\tt not\_all\_equal}, {\tt weightedSum}.\footnote{For a precise definition of these constraints see \cite{GlobalConstraintCatalog}.}

Finally, to give an example of how x4g could be used, we developed a program that takes as input a XCSP instance and, by using x4g, allows us to find a solution by using the deep first strategy natively implemented in Gecode. Since the output of this program follows the output rules of the International Solver Competition, it could be used to let Gecode enter the next International CSP Competition.

\section{Conclusion and Future Work}
\label{conclusions}

In this paper we described x4g, a plug-in that allows the use of Gecode for solving a CSP instance defined using the XCSP format. 
The source code of x4g and of the above mentioned program can be found at \url{http://www.cs.unibo.it/~jmauro/cilc_2011.html}.

This work has to be considered as a first step in the direction of providing a full translation of XCSP constraints into Gecode constraints. As a future work, we would like to support more global constraints and to define efficient ways of decomposing them by using Gecode constraints. We also would like to extend our tool in order to use Gecode for solving the optimization problems that can be defined in XCSP exploiting weighted constraints.

Moreover, we are interested in the development of a FlatZinc/XCSP conveyer that will allow us to add FlatZinc instances into the benchmarks that could be used for comparing Gecode with other constraint solvers. Furthermore, with such a converter it could be also possible to compare the efficiency of our x4g translation with the one of the Gecode/FlatZinc plug-in.

\bibliographystyle{plain}
\bibliography{biblio}

%
%
%

\end{document}